\documentclass[12pt, draftclsnofoot, journal, letter, onecolumn]{IEEEtran} 

\usepackage{graphicx}
\usepackage{epsfig}
\usepackage{latexsym}
\usepackage{amsfonts}
\usepackage{here}
\usepackage{rawfonts}
\usepackage[latin1]{inputenc}
\usepackage[T1]{fontenc}
\usepackage{calc}
\usepackage{url}
\usepackage{enumerate}
\usepackage{color}
\usepackage[tbtags]{amsmath}
\usepackage{amssymb}
\usepackage{upref}
\usepackage{epic,eepic}
\usepackage{times}
\usepackage{dsfont}
\usepackage{comment}
\usepackage{cite}

\newtheorem{theorem}{{\bf Theorem}}
\newtheorem{proposition}{{\bf Proposition}}
\newtheorem{lemma}{{\bf Lemma}}
\newtheorem{corollary}{{\bf Corollary}}

\newcommand{\qed}{\nobreak \ifvmode \relax \else
  \ifdim\lastskip<1.5em \hskip-\lastskip
  \hskip1.5em plus0em minus0.5em \fi \nobreak
  \vrule height0.75em width0.5em depth0.25em\fi}

\newcounter{step}
\newlength{\totlinewidth}
  {\end{list}%
  \rule{\linewidth}{1pt}}
\newcounter{substep}

  {\end{list}}

\newlength{\aligntop}
\newlength{\alignbot}
 \makeatletter
\renewenvironment{align}{%
  \vspace{\aligntop}
  \start@align\@ne\st@rredfalse\m@ne
}{%
  \math@cr \black@\totwidth@
  \egroup
  \ifingather@
    \restorealignstate@
    \egroup
    \nonumber
    \ifnum0=`{\fi\iffalse}\fi
  \else
    $$%
  \fi
  \ignorespacesafterend%
  \vspace{\alignbot}\par\noindent
} \makeatother

\newcommand{\textred}[1]{\textcolor{red}{#1}}
\ifx\noeditingmarks\undefined
  \newcommand{\pgwrapper}[2]{\textred{#1: #2}}
\else
  \newcommand{\pgwrapper}[2]{}
\fi

\IEEEoverridecommandlockouts

\begin{document}

\title{Cognitive Multiple Access Network with Outage Margin in the Primary System\vspace*{-0.3em}}

\author{
\authorblockN{Behrouz Maham, \emph{Member, IEEE}, Petar Popovski, \emph{Senior Member, IEEE}, Xiangyun Zhou \emph{Member, IEEE}, and Are Hj{\o}rungnes, \emph{Senior Member, IEEE}}\\
    \thanks{This work was supported by the Research Council of Norway
    through the project 176773/S10 entitled "Optimized Heterogeneous Multiuser MIMO Networks -- OptiMO" and the project 197565/V30 entitled "Theoretical Foundations of Mobile Flexible Networks -- THEFONE". Behrouz Maham, Xiangyun Zhou, and Are Hj{\o}rungnes are 
    with UNIK -- University Graduate Center, Universitcy of Oslo, Norway. Petar Popovski is with Department of Electronic Systems, Aalborg University, Aalborg, Denmark.
Emails:
\protect\url{{behrouz,arehj,xiangyun}@unik.no,
petarp@es.aau.dk}.}%
}
\maketitle

\begin{abstract}
This paper investigates the problem of spectrally efficient operation of a multiuser uplink cognitive radio system in the presence of a single primary link.
The secondary system applies opportunistic interference cancelation (OIC) and decode the primary signal when such an opportunity is created.
We derive the achievable rate in the secondary system when OIC is used. This scheme has a practical significance, since it enables rate adaptation without requiring any action from the primary system.
The \emph{exact} expressions for outage probability of the primary user are derived, when the primary system is exposed to interference from secondary users. Moreover, approximated formulas and tight lower and upper bounds for the ergodic sum-rate capacity of the secondary network are found.
Next, the power allocation is investigated in the secondary system for maximizing the sum-rate under an outage constraint at the primary system.
We formulate the power optimization problem in various scenarios depending on the availability of channel state information and the type of power constraints, and propose a set of simple solutions.
Finally, the analytical results are confirmed by simulations, indicating both
the accuracy of the analysis, and the fact that the spectral-efficient, low-complexity,
flexible, and high-performing cognitive radio can be designed based on the proposed schemes.
\end{abstract}

\section{Introduction}
Cognitive radio
technology offers efficient use of the
radio spectrum, potentially allowing large amounts of spectrum to
become available for future high bandwidth applications.
A cognitive radio (CR) network (or secondary system) is allowed to use certain
radio resource if it is not causing an adverse interference
to the primary system. Furthermore, the CR should achieve a spectrally
efficient operation under the interference from the primary system.

Some works \cite{dev06,jov09,cheng07} have discussed achievable rates in cognitive radio
from the viewpoint of information theory. The seminal
work \cite{jov09} on the achievable rate of a single cognitive radio
user considers the constraints that there is no interference
to the primary user, and the primary encoder-decoder pair
is oblivious of the presence of cognitive radios. References \cite{cheng07,han08}
extend the results of \cite{jov09} to multiple cognitive radio users and
characterize the cognitive radio's achievable rate region for
Gaussian multiple-access channels (MACs). Maximization of
the cognitive radio's sum-rate on Gaussian MAC then raises
the problem of the allocation of each cognitive user's power
ratio \cite{han08}.
In \cite{han09,han10}, two spectrum sharing protocols based on cooperative relay transmission are proposed. In particular, \cite{han10} considers a spectrum access protocol with multiple CRs. Furthermore, the problem of power allocation in CR networks has been considered in a number of recent works. For example, in \cite{hoa10}, the authors proposed some mixed distributed-centralized power control for multiuser CR to maximize the total throughput while maintaining a required signal to interference plus noise ratio (SINR) for primary users. However, in contrast to our work, they assumed that CR users cannot transmit simultaneously on one frequency band. In \cite{gao09}, an energy constrained wireless CR ad hoc network is considered, where each node is equipped with CR and has limited battery energy. Given the data rate requirement and maximal power limit, a constrained optimization problem is formulated in \cite{gao09} to minimize the energy consumption, while avoid introducing interference to the existing users. A power control scheme for maximum sum-rate of fading multiple access network is proposed in \cite{wan10} under instantaneous interference power constraint at the primary network. In \cite{kan09}, with perfect
channel state information (CSI) on the channels from the secondary user
transmitter to the secondary and primary receivers, the optimal power
allocation strategies to achieve the ergodic/outage capacities of
a single secondary user fading channel subject to both secondary user's transmit and
interference power constraints were studied.

As mentioned in \cite{set09}, there are two types of interference
in the system due to the coexistence of primary users and secondary users. One is introduced by primary users into the secondary users bands, and the other is introduced by the secondary users into the primary users'
bands. Peaceful coexistence of secondary users with primary users requires that the
secondary interference at a primary receiver is below a certain threshold \cite{set09}.
The primary should operate with a certain \emph{margin}, which allows to accommodate
transmissions in the secondary system without degrading the target performance of the primary. The margin can take several forms:
(a) Time - the primary communicates less than $100\%$ of
the time; (b) frequency - the primary is using only part of
its allocated spectrum; or (c) interference - the secondary can
transmit by keeping the interference below some threshold \cite{pop09}.
The secondary needs to perform spectrum sensing and identify
its transmission opportunity, which in the cases (a) and (b)
consists of detecting the spectrum hole \cite{hay05}, while in (c) it
detects the interference induced to the primary receivers \cite{zhang08}.
Here, we consider scenarios that deal with the interference
margin by keeping the outage probability or signal to interference plus noise ratio (SINR) in the primary system at
an acceptable value.

Moreover, we investigate the problem of spectrally efficient
operation in a multiuser secondary under interference from a primary system. The primary system adapts its data rate
for the primary terminals and the chosen primary transmission rate is independent of the SNR at which the primary signal is
received at the secondary receiver.
Upon a simultaneous reception of a secondary signal and a primary signal,
a secondary receiver observes a multiple access channel. The objective of the
secondary receiver is to decode the primary signal only to help to achieve
a better secondary rate; the secondary receiver is not interested in the primary data. The authors in \cite{pop07}
call this opportunistic interference cancelation (OIC), as the
decodability of the primary system signal at the secondary receiver depends on the
opportunity created by the selection of the data rate in the primary system
and the SNR on the link between the primary transmitter and the secondary receiver.
In this paper, we extend the result in \cite{pop07} from single user secondary system to uplink multiuser secondary network. Hence, the secondary receiver observes a MAC of two group of users: The desired secondary multiuser transmitters and the undesired primary transmitter.

Our main contributions can be summarized
as follows: 
\begin{enumerate}
%
%
\item This paper considers efficient resource allocation for sum--rate maximization of the secondary rates over a Gaussian MAC. We extend the OIC to the case of multiuser secondary network, and depending on decodability of primary signal at the secondary receiver and channel conditions, appropriate rates can be assigned to secondary users.
%
%
\item 
%
We derive closed-from expressions for the outage probability at the primary user when there are multiple secondary interferers.
The simplicity of the derived expressions can give insight on performance of the system and lead to system optimization.
%
\item
A set of ergodic sum-rate capacity bounds and approximations are derived in secondary with rate adaptation using OIC scheme. The numerical results verify the tightness of the bounds.
%
%
\item We formulate the problem of maximizing the secondary uplink sum-rate capacity for an outage--restricted primary system under different
assumptions about the CSI knowledge at the secondary users.
    We propose simple power control schemes to maximize the secondary uplink capacity
given the outage probability constraint.
    The proposed system can achieve
considerable increase in spectrum-efficiency compared to orthogonal transmission strategies.
%
%
%
\end{enumerate}

%
%
The remainder of this paper is organized as follows:
In Section II, the system model and protocol description are given.
A spectrally efficient operation for CR is studied in Section~III.
The closed--form expressions for some performance metrics are presented in Section IV, which are utilized for optimizing the system.
Section V presents the problem of maximization of the secondary capacity through power control of the secondary devices and under interference constraints at the primary system.
%
In Section VI, the overall system performance is presented for
different numbers of users and channel conditions, and the correctness of the analytical
formulas is confirmed by simulation results.
Conclusions are presented in Section VII.

%
\emph{Notations}: The superscripts $(\cdot)^t$, $(\cdot)^H$, and $(\cdot)^*$ stand for
transposition, conjugate transposition, and element-wise
conjugation, respectively.
The expectation operation is denoted by $\mathbb{E}\{\cdot\}$.
%
%
The symbol $|x|$
is the absolute value of the scalar $x$, while
$[x]^+$ denotes $\max\{x,0\}$.
The logarithms $\log_2$ and $\log$ are the based two logarithm and the natural logarithm, respectively.

\section{System Model and Protocol Description}
\begin{figure}[e]
  \centering
  \includegraphics[width=4.5in]{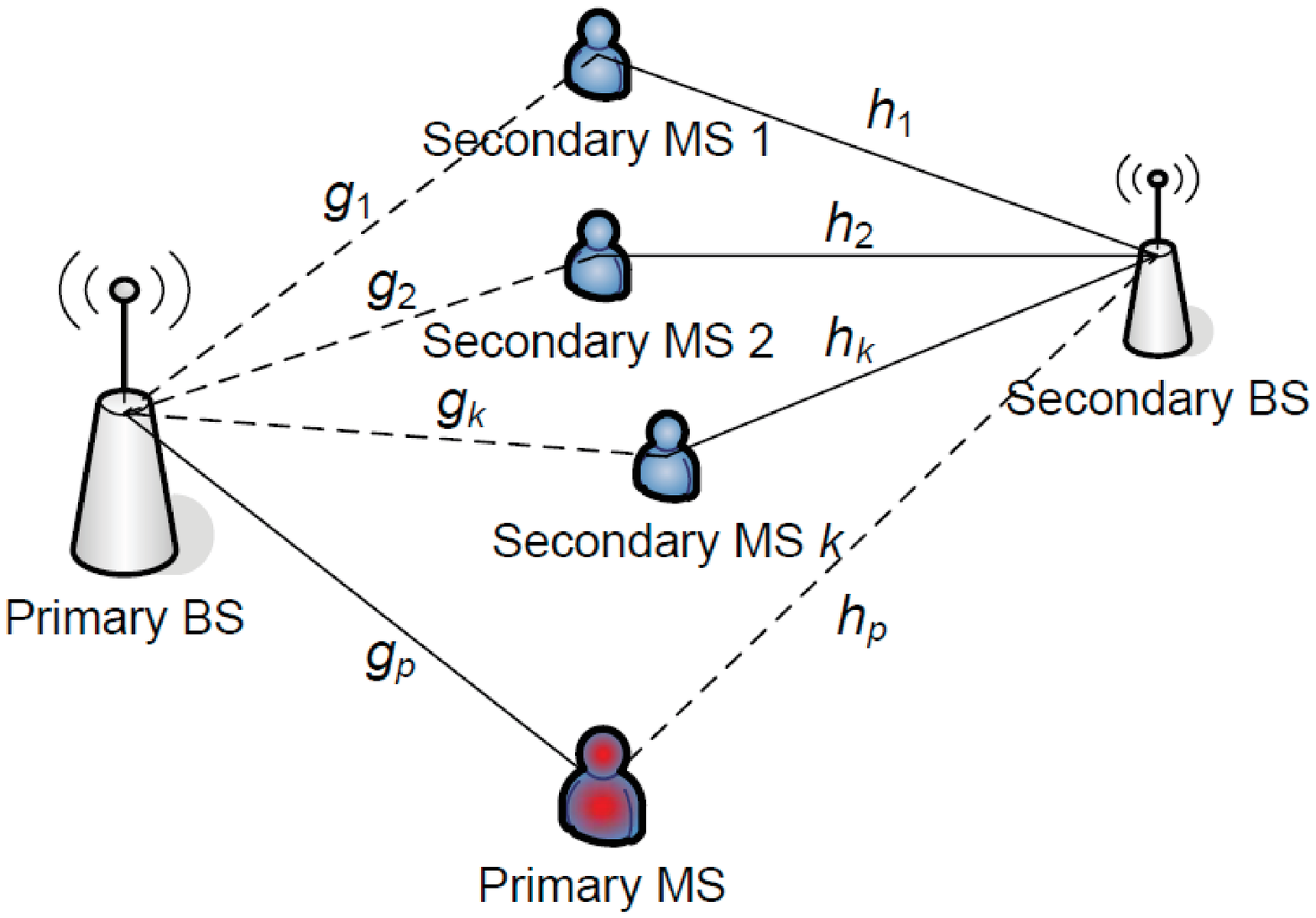}\\
  \caption{Wireless network with multiple cognitive users access.
  }\label{f0}
\end{figure}

We consider the scenario depicted on Fig.~\ref{f0}, consisting of a primary transmitter, a primary receiver, $K$ secondary transmitters and one secondary receiver. All the nodes are equipped with a single antenna.
In this model, a
primary mobile station (MS) is communicating with the primary
base station (BS) and there are multiple secondary MS. The secondary MS desire to
access to secondary BS using primary frequencies without
license. It is assumed that $g_p$ is the channel coefficient from primary MS to
primary BS, and $g_k$, $k = 1, 2, \ldots,K$, is the channel coefficient of
the interference link from secondary MS $k$ to the primary BS. In addition, $h_k$, $k = 1, 2, \ldots,K$, is the channel coefficient from MS $k$ to the secondary BS and $h_p$ is the interference link from the primary MS to secondary BS. Throughout this paper, we assume that all channels are modeled as independent Rayleigh fading, and the primary and secondary receivers have additive white Gaussian noise with variance $\mathcal{N}_p$ and $\mathcal{N}_s$, respectively.
The average power of the primary user is $P_0$ and the average power of secondary user $k$ is assumed to be $P_k$, $k = 1, 2, \ldots,K$, respectively.

\subsection{Primary System}
The primary MS uses fixed transmission rate $R_p$ in the uplink.
In absence of interference, the signal
received at the primary BS is given by
\begin{equation}\label{2}
    {y}_{p}=\sqrt{P_{0}}\,g_{p}x_p+{v}_{p},
\end{equation}
where $x_p$ is the signal sent by the primary user, normalized as $\mathbb{E}\{|x_p|^2\}=1$, $v_p$ is the additive Gaussian noise at the primary BS with variance $\mathcal{N}_p$, and $P_0$ is the transmit power from the primary MS.
Considering normalized bandwidth, the achievable instantaneous rate is $\log_2\left(1+\frac{P_0|g_p|^2}{\mathcal{N}_p}\right)$.

The minimum SNR to support rate $R_p$ is denoted by $\gamma_{\text{th}} = 2^{R_p}-1$. If the achievable rate is lower than $R_p$, then outage occurs.
Let $\rho_m$ be the maximal allowed outage probability at the primary receiver.
If $\rho_m>\rho_0$, where $\rho_0$ is the outage probability in absence of secondary interference, then the receiver has an \emph{outage margin} and additional interference can be received from the secondary transmission without violating the target operation regime of the primary system.
Thus, in presence of interference, the interfered signal at the primary receiver can be represented as
\begin{equation}\label{3}
    {y}_{p}=\sqrt{P_{0}}\,g_{p}\,x_p+\sum_{k=1}^{K}\sqrt{P_{k}}\,g_{k}\,x_{k}+{v}_{p},
\end{equation}
where $P_k$ and $x_k$ are the allocated power and the transmit signal of secondary MS $k$, respectively.
For primary user's receiver, its data rate is obtained by treating the secondary users as noise:
\begin{equation}\label{4}
    r_{p}=\log_2\left(1+\frac{P_0|g_p|^2}{\mathcal{N}_p+\sum_{k=1}^{K}P_k|g_k|^2}\right).
\end{equation}

\subsection{The Secondary System}
The secondary system consists of $K$ users accessing the same secondary BS. We consider a multiuser space-division multiple-access (SDMA)-based cognitive radio system, which assumes that multiple mobiles simultaneously transmit data streams on the same resource (frequency and time).
For uplink SDMA, collaborative spatial multiplexing (CSM), which usually considers mobile stations with one transmission antenna, is a very efficient scheme increasing the uplink throughput compared to orthogonal transmission schemes. It was adopted for uplink SDMA scheme in IEEE 802.16 systems \cite{ieee16}. Due to the broadcast nature of wireless channels, the capacity analysis of this scheme becomes equivalent to information-theoretic transmission strategy of superposition coding \cite{tse05}.

The received signal at the secondary BS is given as
\begin{equation}\label{300}
    {y}_{s}=\sum_{k=1}^{K}\sqrt{P_{k}}\,h_{k}\,x_{k}+\sqrt{P_{0}}\,h_{p}\,x_p+{v}_{s},
\end{equation}
where ${v}_{s}$ is the Gaussian noise at the secondary BS with variance $\mathcal{N}_s$.
We assume that the signal transmitted from the $k$-th secondary user is $\sqrt{P_{k}}\,x_{k}$, where $\mathbb{E}\{|x_k|^2\}=1$, for $k=1,2,\ldots,K$.
The optimal uplink capacity is achieved by superposition coding at the secondary users and successive interference cancelation (SIC) or generalized decision feedback equalizer~(GDFE) at the secondary BS \cite{tse05}.

\subsection{Channel Knowledge Requirement and Estimation}
The estimation of the instantaneous channel gains of the primary interference link $h_p$, the primary link $g_p$, and the secondary interference links $g_k$, $k=1,\ldots,K$, might not be feasible for secondary users. Thus, here we consider two cases. It is assumed that only the interference channels statistics, i.e., $\sigma^2_{h_p}=\mathbb{E}\{|h_p|^2\}$ and $\sigma^2_{g_k}=\mathbb{E}\{|g_k|^2\}$, $k=1,\ldots,K$ are known at the secondary MS.
The value of $\sigma^2_{g_k}$, $k=1,\ldots,K$ can be inferred by
listening to the downlink transmissions of primary system. On the other
hand, the determination of
$\sigma^2_{g_p}=\mathbb{E}\{|g_p|^2\}$ requires either explicit signaling from the primary system to the secondary users or that secondary users know the location of the primary MS or another indirect way of knowing. Such an indirect way can be achieved
by having the secondary MS overhear the transmissions of the primary MS and based on the ACK/NACK
sent by the primary BS, assess the outage probability at the primary BS in the absence of interference. This value of the outage probability has a one-to-one correspondence with $\sigma^2_{g_p}$.

For the CSI knowledge of the secondary uplink channels at the transmitters, we consider two scenarios. In the first scenario, we assume that only statistics of cognitive uplink channels, i.e., $\sigma^2_{h_k}$, $k=1,\ldots,K$, are known by the secondary users. Thus, ergodic capacity is used as performance metric for power optimization.
In the second scenario, it is assumed that instantaneous channel magnitude of $|h_k|$ is available at the secondary users, and thus, sum-rate capacity of the secondary system in \eqref{1b} and \eqref{1x} can be maximized to find the optimal transmit power.

\begin{figure}[e]
  \centering
  \includegraphics[width=3.2in]{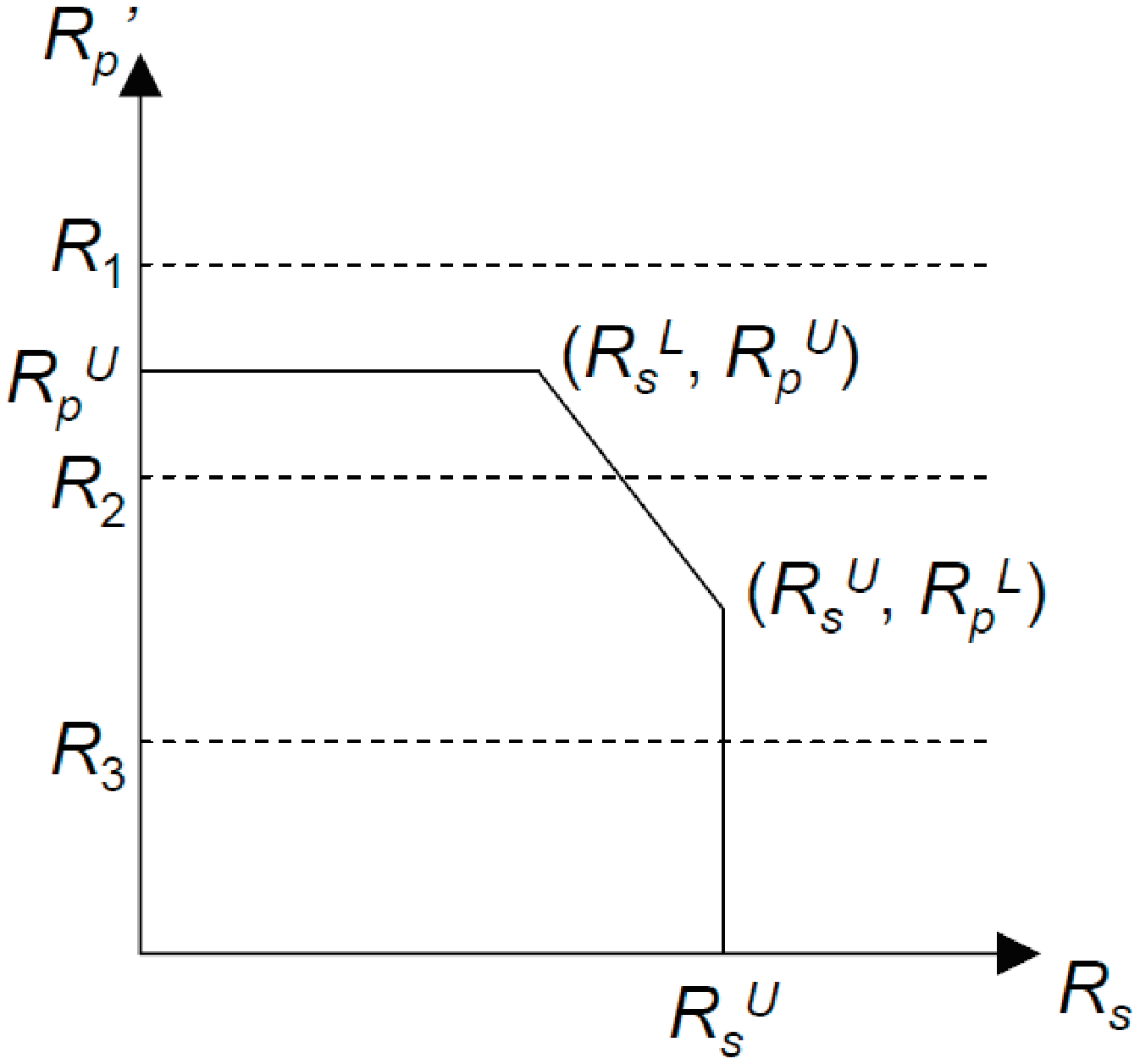}\\
  \caption{The region of achievable rate pair ($R_s$, $R_p'$) of secondary system sum rate and primary rate from secondary receiver viewpoint.
  }\label{fp}
\end{figure}
\section{Opportunistic Interference Cancelation in Cognitive MAC}
The concept of OIC is introduced in \cite{pop07}. However, \cite{pop07} considered the \emph{single} secondary user. In this section, we generalize this to the case of multiuser secondary network. 
Using OIC, the interference from the primary transmitter is canceled whenever such an opportunity is created by (a) selection of the data rate in the primary system $R_p$ and (b) the link quality between the primary transmitter and the secondary receiver, i.e., $h_p$.
Considering the co-existence of primary system with secondary system, the cognitive MAC can be regarded as a Gaussian MAC with common interference.
Define $R_s$ and $R_p'$ bits/s/Hz as the total bandwidth-normalized transmission rate of the uplink multiuser secondary and the achievable rate of the primary signal at the secondary BS,
respectively. Note that the actual primary user transmission rate $R_p$ is fixed and could be different from $R_p'$.
The secondary receiver can reliably decode both the primary and secondary signals if the rates $R_p'$ and $R_s$ are within the capacity region of the multiple access channel (Fig.~\ref{fp}):
\begin{align}\label{1x}
    R_s&\leq C\left(\sum_{k=1}^{K}\frac{P_k|h_k|^2}{\mathcal{N}_{s}}\right)\triangleq R_s^U,\,\,
    \nonumber\\
    R_p'&\leq C\left(\frac{P_0|h_p|^2}{\mathcal{N}_{s}}\right)\triangleq R_p^U,
    \nonumber\\
    R_s+R_p'&\leq C\left(\frac{P_0|h_p|^2}{\mathcal{N}_{s}}+\sum_{k=1}^{K}\frac{P_k|h_k|^2}{\mathcal{N}_{s}}\right),
\end{align}
where $C(x)=\log_2(1+x)$.

We assume that $R_p$ is given a priori at the secondary receiver.
Now, we determine the maximal achievable rate $R_s$. In absence of the primary signal, we have
$$R_s=C\left(\sum_{k=1}^{K}\frac{P_k|h_k|^2}{\mathcal{N}_{s}}\right).$$
Using OIC, the cognitive radio makes the best possible use of the knowledge about the primary system. In order to determine the maximum achievable rate, two regions for $|h_p|^2$ are considered.

\emph{Weak Interference}:
When $|h_p|^2<\frac{\mathcal{N}_{s}}{P_0}(2^{R_p}-1)$, the secondary BS cannot decode the primary signal and we have
\begin{align}\label{1x}
    R_s= C\left(\sum_{k=1}^{K}\frac{P_k|h_k|^2}{\mathcal{N}_{s}+P_0|h_p|^2}\right)=
    \log_2\left(1+\frac{\sum_{k=1}^{K}P_k|h_k|^2}{\mathcal{N}_{s}+P_0|h_p|^2}\right)\triangleq R_s^L.
\end{align}
This is equivalent to the case that the maximal decodeable rate $R^U_p$ should be less than the
actual primary rate $R_p$,
where $R_p$ is depicted as a constant $R_1$ in Fig.~\ref{fp}.
Thus, when the primary signal is not strong, it is treated as a noise at the secondary receiver, and the sum--rate is given by \eqref{1x}.

In the region $|h_p|^2\geq\frac{\mathcal{N}_{s}}{P_0}(2^{R_p}-1)$, the secondary receiver can decode the primary
signal and $R_s$ is chosen such that ($R_s$, $R_p'$) belongs to the
achievable rate region, determined for the given channel gains. When $|h_p|^2\geq\frac{\mathcal{N}_{s}}{P_0}(2^{R_p}-1)$, or equivalently, $R_p^U\geq R_p$, we have two cases.
\\
\emph{Medium Interference}:
If $R_p^L<R_p$ where
\begin{align}\label{1c}
    R_p^L\triangleq C\left(\frac{P_0|h_p|^2}{\mathcal{N}_{s}+\sum_{k=1}^{K}P_k|h_k|^2}\right),
\end{align}
the achievable rate is chosen from the segment between the corner points ($R_s^L$, $R_p^U$) and ($R_s^U$, $R_p^L$) in Fig.~\ref{fp}. In this case, the value of $R_p$ can be set as $R_2$ shown in Fig.~\ref{fp}, where $R_2$ is a positive constant.
In other words,
\begin{align}\label{2c}
    \frac{\mathcal{N}_{s}}{P_0}(2^{R_p}-1)\leq|h_p|^2<
    \frac{2^{R_p}-1}{P_0}\left(\mathcal{N}_{s}+\sum_{k=1}^{K}P_k|h_k|^2\right).
\end{align}
For this case, observing Fig.~\ref{fp}, the achievable rate for the secondary system can be calculated as
\begin{align}\label{2z}
    R_s&= C\left(2^{-R_p}\left[\frac{P_0|h_p|^2}{\mathcal{N}_{s}}+
    \sum_{k=1}^{K}\frac{P_k|h_k|^2}{\mathcal{N}_{s}}-2^{R_p}+1\right]\right)
    \nonumber\\
    &=-R_p+\log_2\left(\frac{P_0|h_p|^2}{\mathcal{N}_{s}}+
    \sum_{k=1}^{K}\frac{P_k|h_k|^2}{\mathcal{N}_{s}}+1\right).
\end{align}

\emph{Strong Interference}:
Another scenario is when $R_p^L \geq R_p$ where we have a strong interference from the primary system. In this case, the maximum achievable rate is chosen from the vertical segment in Fig.~\ref{fp}. In this case, the value of $R_p$ can be set as $R_3$ shown in Fig.~\ref{fp}, where $R_3$ is a positive constant. In other words,
\begin{align}\label{3c}
    |h_p|^2\geq
    \frac{2^{R_p}-1}{P_0}\left(\mathcal{N}_{s}+\sum_{k=1}^{K}P_k|h_k|^2\right).
\end{align}
For this case, the achievable rate for the secondary system can be calculated as
\begin{align}\label{1b}
    R_s=C\left(\sum_{k=1}^{K}\frac{P_k|h_k|^2}{\mathcal{N}_{s}}\right)
    =\log_2\left(1+\frac{\sum_{k=1}^{K}P_k|h_k|^2}{\mathcal{N}_{s}}\right).
\end{align}
Thus, the maximal achievable rate in the secondary system is obtained whenever the primary signal is decodable and the condition in \eqref{3c} is fulfilled. In other words, when the interference from the primary sender is strong, and the secondary receiver is able to decode and remove the interference from the primary transmitter, the achievable rate is given by \eqref{1b}.
Note that when there is cooperation between the primary and secondary transmitters, we can achieve so-called "clean-MAC" capacity as \eqref{1b} for all interference conditions (see e.g., \cite{han08} and \cite{kim04}).
Since it is hard to realize the case of cooperation with cognitive MAC which requires a substantial amount of the data exchange, we assume there is no cooperation in a sense of data exchange between primary and secondary systems.
A less optimal strategy would be to treat the primary signal an undecodable interference, even when interference is strong.

\section{Performance Metrics}
\subsection{Outage Probability of Primary System with Interference Margin}
As stated above, the interference from the secondary users should be kept below a threshold in order to coexist with the primary system.
Thus, the secondary system should choose the power $P_k$, $k=1,2,\ldots,K$, in such a way that the outage performance for the primary system is not violated.

In the following, the outage probability $\rho_{\text{out}} \triangleq \text{Pr}\{r_p < R_p\}$ of the primary BS is
investigated, which describes the probability that the transmit rate
$R_p$ is greater than the supported rate $r_p$ in \eqref{4}. This probability which is
expressed as a cumulative distribution function~(CDF) depends on
the fixed transmission parameters and the channel condition within
the primary system and the secondary cognitive network. By defining $\gamma_{\text{th}}\triangleq(2^{R_p}-1)$, the outage probability at the primary user can be represented as
\begin{equation}\label{6}
    \rho_{\text{out}} =
\text{Pr}\left\{\frac{P_0|g_p|^2}{\mathcal{N}_p+\sum_{k=1}^{K}P_k|g_k|^2
}<\gamma_{\text{th}}\right\}.
\end{equation}

\begin{proposition}
Consider a finite set of independent random variables $X$ and $\mathcal{Y}=\{Y_1,\ldots,Y_K\}$, with exponential distribution and non-identical mean of $\sigma_{x}^2$ and $\sigma_{k}^2$, $k=1,\ldots,K$, respectively.
The CDF of the signal-to-noise ratio
\begin{equation*}\label{7}
\text{SINR}=\frac{X}{1+\sum_{k=1}^{K}Y_k},
\end{equation*}
can be calculated as
\begin{equation}\label{8}
    \text{Pr}\left\{\text{SINR}<\gamma\right\} = 1- e^{-\frac{\gamma}{\sigma_{x}^2}}\prod_{k=1}^{K}\left(1+\frac{\sigma_{k}^2}{\sigma_{x}^2}
    \gamma\right)^{-1}.
\end{equation}

\end{proposition}
\begin{proof}
By marginalizing over the set of independent random variables $\mathcal{Y}$, the CDF of the SINR can be calculated as
\begin{align}\label{9}
    \text{Pr}\left\{\text{SINR}<\gamma\right\}&=
    \int_{0;K-\text{fold}}^{\infty}\text{Pr}\left\{X<\gamma+\gamma\sum_{k=1}^{K}y_k\right\}\,\prod_{k=1}^{K}p_k(y_k)\,dy_k
    \nonumber\\
    &=1-\int_{0;K-\text{fold}}^{\infty}e^{-\frac{\gamma(1+\sum_{k=1}^{K}y_k)}{\sigma_{x}^2}}\,
    \prod_{k=1}^{K}\frac{e^{-\frac{y_k}{\sigma_{k}^2}}}{\sigma_{k}^2}\,dy_k.
\end{align}
By solving the integrals is the second equation of \eqref{9}, the CDF is obtained as \eqref{8}.
\end{proof}

From Proposition 1 and by defining $X=\frac{P_0|g_p|^2}{\mathcal{N}_p}$ and $Y=\frac{P_k|h_k|^2}{\mathcal{N}_p}$, the outage probability in \eqref{6} can be written as
\begin{equation}\label{10}
    \rho_{\text{out}} = 1- e^{-\frac{\gamma_{\text{th}}\,\mathcal{N}_p}{P_0\sigma_{g_p}^2}}\prod_{k=1}^{K}
    \left(1+\frac{P_k\,\sigma_{g_k}^2}{P_0\,\sigma_{g_p}^2}
    \gamma_{\text{th}}\right)^{-1},
\end{equation}
where $\sigma_{g_p}^2$ and $\sigma_{g_k}^2$, $k=1,\ldots,K$, are the mean of the channel coefficients $g_p$ and $g_k$, $k=1,\ldots,K$, respectively.


\subsection{Ergodic Capacity of Cognitive Multiple Access Channel}
For the ergodic sum-rate performance given as $\overline{R}_{s}=\mathbb{E}\{R_{s}\}$, where $\mathbb{E}\{\cdot\}$ denotes the expectation operation, we have from \eqref{1b}
\begin{equation}\label{11}
    \overline{R}_{s}=\mathbb{E}\left\{\log_2\left(1+\frac{\sum_{k=1}^{K}P_k|h_k|^2}{\mathcal{N}_s}\right)\right\}.
\end{equation}

\subsubsection{Upper-Bound}
By the fact that $\log_2(1+x)$ is a concave function, we derive an upper-bound for the ergodic capacity of the secondary system.
In order to derive a upper-bound on the above expression, we use Jensen's inequality
\begin{align}\label{12}
    \overline{R}_{s}&\leq
    \log_2\left(1+\frac{\mathbb{E}\left\{\sum_{k=1}^{K}P_k|h_k|^2\right\}}{\mathcal{N}_s}\right)
=\log_2\left(1+\frac{\sum_{k=1}^{K}P_k\sigma_{h_k}^2}{\mathcal{N}_s}\right).
\end{align}
Similarcly, in the case of the medium received primary SNR at the secondary receiver, i.e., when the condition in \eqref{2c} is satisfied, an upper-bound for ergodic capacity of \eqref{2z} can be written as
\begin{align}\label{12q}
    \overline{R}_{s}&\leq
-R_p+\log_2\left(1+\frac{P_0\sigma_{h_p}^2}{\mathcal{N}_s}+
\frac{\sum_{k=1}^{K}P_k\sigma_{h_k}^2}{\mathcal{N}_s}\right).
\end{align}

\subsubsection{Lower-Bound}
A lower-bound on the ergodic capacity in \eqref{11} can be calculated by the fact that $\log_2(1+a\,e^x)$ is a convex function with $a>0$. Thus, applying Jensen's inequality, we have
\begin{align}\label{12b}
    \overline{R}_{s}&\geq
    \log_2\left(1+\frac{\exp\left(\mathbb{E}\left\{\log\left[\sum_{k=1}^{K}P_k|h_k|^2\right]\right\}\right)}
    {\mathcal{N}_s}\right).
\end{align}

Assuming that secondary users have the same distance to the secondary BS, i.e., $|h_k|^2$ are i.i.d. random variables, a closed-form solution for the expression in \eqref{12b} is given by
\begin{align}\label{12c}
    \overline{R}_{s}&\geq
    \log_2\left(1+\frac{P_s\,\sigma_{h}^2}{\mathcal{N}_s}\exp\left(\sum_{k=1}^{K-1}\frac{1}{k}-\kappa\right)\right)
\end{align}
where $\kappa\approx 0.577$ is Euler's constant, $P_k=P_s$, and $\sigma_{h_k}^2=\sigma_{h}^2$, $k=1,\ldots,K$. The result in \eqref{12c} is obtained by applying the techniques in \cite{oym03}
and the fact that for no CSI at the transmitters, the ergodic sum capacity of
a $K$ users MAC channel, where each user has a single transmit
antenna, is equivalent to the ergodic capacity of a single-user
system with $K$ transmit antennas \cite[Proposition 1]{rhe03}.

Now we consider the case of non-i.i.d. random variables $|h_k|^2$, $k=1,\ldots,K$. Define the vector $[x_1,\ldots,x_K]$ of multiple
variables. Then,
$\log_2(1+\sum_{k=1}^{K}a_k\,e^{x_k})$ is a convex function on $\mathds{R}^K$ for arbitrary $a_k>0$ (see e.g. \cite[Lemma 3]{zhan05}). Thus, applying Jensen's inequality in \eqref{11}, we have
\begin{align}\label{120b}
    \overline{R}_{s}&\geq
    \log_2\left(1+\sum_{k=1}^{K}\frac{P_k}
    {\mathcal{N}_s}\exp\left(\mathbb{E}\left\{\log\left[|h_k|^2\right]\right\}\right)\right).
\end{align}
From \cite{zhan05}, we know that $\mathbb{E}\left\{\log\left[|h_k|^2\right]\right\}=\log(\sigma_{h_k}^2)+\psi(1)=\log(\sigma_{h_k}^2)-\kappa$ where $\psi(\cdot)$ is the digamma or psi function \cite[Eq. (8.360)]{gra96}.
Thus, a closed-form solution for the expression in \eqref{120b} is given by
\begin{align}\label{120c}
    \overline{R}_{s}&\geq
    \log_2\left(1+\sum_{k=1}^{K}\frac{P_k\sigma_{h_k}^2}{\mathcal{N}_s}\exp(-\kappa)\right).
\end{align}
Similarly, in the case of the medium received primary SNR at the secondary receiver, i.e., when the condition in \eqref{2c} is satisfied, a lower-bound for ergodic capacity of \eqref{2z} can be written as
\begin{align}\label{12q}
\overline{R}_{s}&\geq-R_p+\log_2\left(1+\frac{P_0\sigma_{h_p}^2}{\mathcal{N}_s}\exp(-\kappa)+
\frac{\sum_{k=1}^{K}P_k\sigma_{h_k}^2}{\mathcal{N}_s}\exp(-\kappa)\right).
\end{align}

\subsubsection{Ergodic Capacity of Cognitive Network with Weak Interference}
Now, we investigate ergodic capacity for the case of weak interference from primary user to the secondary receiver. From \eqref{1x}, an upper-bound for the ergodic capacity of the secondary system is given by
\vspace{.2cm}
\begin{align}\label{12ss}
    \overline{R}_{s}&
    =\mathbb{E}_{|h_p|^2<\,c_p}\!\left\{\log_2\!\left(1+\frac{\sum_{k=1}^{K}P_k|h_k|^2}{\mathcal{N}_s+P_0|h_p|^2}\right)
    \right\}
    \leq\log_2\!\left(1+\mathbb{E}_{|h_p|^2<\,c_p}
    \!\left\{\frac{\sum_{k=1}^{K}P_k|h_k|^2}{\mathcal{N}_s+P_0|h_p|^2}\right\}\right)
    \nonumber\\
    &\leq\log_2\!\left(1+\frac{\sum_{k=1}^{K}P_k\sigma_{h_k}^2}{\mathcal{N}_s+P_0\mathbb{E}_{|h_p|^2<\,c_p}\!
    \left\{|h_p|^2\right\}}\right)
    =\log_2\!\left(\!1\!+\frac{\sum_{k=1}^{K}P_k\sigma_{h_k}^2}{\mathcal{N}_s+P_0\sigma_{h_p}^2(1-e^{
    -\frac{c_p}{\sigma_{h_p}^2}})-P_0 \, c_p\, e^{
    -\frac{c_p}{\sigma_{h_p}^2}}}\right),
\end{align}
where $c_p=\frac{\mathcal{N}_{s}}{P_0}(2^{R_p}-1)$ and in the two inequalities above we used Jensen's inequality.
Similar to \eqref{120b}, an upper-bound for $\overline{R}_{s}$ in this case is obtained as
\begin{align}\label{140b}
    \overline{R}_{s}&\geq
    \log_2\left(1+\sum_{k=1}^{K}P_k\exp\left(\mathbb{E}\left\{\log\left[|h_k|^2\right]\right\}
    -\mathbb{E}_{|h_p|^2<\,c_p}\left\{\log\left[\mathcal{N}_s+P_0|h_p|^2\right]\right\}\right)\right).
\end{align}
Since $\mathbb{E}_{|h_p|^2<\,c_p}\left\{\log\left[\mathcal{N}_s+P_0|h_p|^2\right]\right\}\leq\log\left[\mathcal{N}_s+P_0
\mathbb{E}_{|h_p|^2<\,c_p}\left\{|h_p|^2\right\}\right]$, a close-form lower-bound for \eqref{140b} can be written as
\vspace{.1cm}
\begin{align}\label{12x}
\overline{R}_{s}&    \geq\log_2\!\left(1+\frac{\sum_{k=1}^{K}P_k\sigma_{h_k}^2}{
\mathcal{N}_s+P_0\sigma_{h_p}^2(1-e^{
    -\frac{c_p}{\sigma_{h_p}^2}})-P_0 \, c_p\, e^{
    -\frac{c_p}{\sigma_{h_p}^2}}}\exp(-\kappa)\right).
\end{align}
Furthermore, if the secondary links $h_k$ have i.i.d. distribution, a tighter lower-bound can be obtained using the bound in \eqref{12c} as
\begin{align}\label{11oo}
    \overline{R}_{s}&\geq\log_2\left(1+\frac{P_s\,\sigma_{h}^2}{\mathcal{N}_s+P_0\sigma_{h_p}^2(1-e^{
    -\frac{c_p}{\sigma_{h_p}^2}})-P_0 \, c_p\, e^{
    -\frac{c_p}{\sigma_{h_p}^2}}}
    \exp\left(\sum_{k=1}^{K-1}\frac{1}{k}-\kappa\right)\right).
\end{align}

\section{Permissible Power Allocation on Gaussian Cognitive MAC}
In this section, permissible power levels in the secondary system are investigated. First, we derive the power allocation for the case that the secondary user experiences strong interference from the primary sender and interference is decoded. Next, we show that for the case of weak interference and treating interference as noise, the same power allocation schemes can be applied.

\subsection{Power Optimization with Known Cognitive MAC Statistical CSI at Secondary Users}
Here, we assume that instantaneous CSI of cognitive multiple access channel gains are not available at the secondary users. However, it is assumed that the statistics of the secondary channels, i.e., $\sigma_{h_k}^2$, $k=1,\ldots,K$, and interference channels $\sigma_{h_k}^2$, $k=1,\ldots,K$, should be estimated for calculating the power control coefficients.
Therefore, we consider the ergodic capacity as a performance metric for the cognitive MAC system.

Before formulating the problem of maximizing the rate given the outage constraint, we present the following lemma:
\begin{lemma}
The optimum point for maximizing the sum-rate
capacity of cognitive MAC using OIC over the feasible set of the power coefficients $P_k$, $k=1,\ldots,K$, is same as maximizing the rate given in \eqref{1b}, i.e., clean-MAC capacity.
\end{lemma}
\begin{proof}
By defining $\gamma_k=\frac{|h_k|^2}{\mathcal{N}_{s}}$, $\gamma_p=\frac{P_0|h_p|^2}{\mathcal{N}_{s}}$, and combining \eqref{1x}, \eqref{2z}, and \eqref{1b}, the sum-rate capacity at the secondary receiver is given by
\begin{align}\label{13q}
C_{\text{sum}}\left(\mathcal{P},\{\gamma_k\}_{k=1}^{K},\gamma_p, R_p\right)=\left\{\begin{array}{cc}
                                                \log_2\left(1+\frac{\Psi_{\mathcal{P}}}{1+\gamma_p}\right), & \text{if}\,\,\gamma_p<\alpha,
\\
-R_p+\log_2\left(1+\gamma_p+
    \Psi_{\mathcal{P}}\right), & \text{if}\,\, \alpha\leq\gamma_p<
    \alpha\left(1+\Psi_{\mathcal{P}}\right),\\
    \log_2\left(1+\Psi_{\mathcal{P}}\right), & \text{if}\,\, \gamma_p\geq
    \alpha\left(1+\Psi_{\mathcal{P}}\right).
                                              \end{array}
    \right.
\end{align}
where $\alpha=2^{R_p}-1$, $\Psi_{\mathcal{P}}=\sum_{k=1}^{K}P_k\gamma_k$, and
$$\mathcal{P}=\left\{P_k, k=1,\ldots,K: 1- e^{-\frac{\gamma_{\text{th}}\,\mathcal{N}_p}{P_0\sigma_{g_p}^2}}\prod_{k=1}^{K}
    \left(1+\frac{P_k\,\sigma_{g_k}^2}{P_0\,\sigma_{g_p}^2}
    \gamma_{\text{th}}\right)^{-1}\leq\rho_{m}, \,\,P_k\geq 0, \forall k\right\}.$$
As it can be seen from \eqref{13q}, for a given primary parameters, i.e., $R_p$, $P_0$, and $|h_p|^2$, $C_{\text{sum}}$ is an increasing function of $\Psi_{\mathcal{P}}$. Moreover, $\Psi_{\mathcal{P}}$ is weighted sum of the power coefficients $P_k\in \mathcal{P}$ with non-negative weights. Hence, the optimum power coefficients $P_k^*$, $k=1,\ldots,K$, for maximizing the strong interference capacity, i.e., $\log_2\left(1+\Psi_{\mathcal{P}}\right)$ is the same as the optimum power coefficients for maximizing $C_{\text{sum}}$.
\end{proof}

Now, using Lemma 1, we formulate the problem of power allocation in cognitive multiple access channel (or uplink cognitive network).
As stated in the previous section, the performance metric for network optimization is the ergodic capacity, or more precisely, its lower bound
\eqref{120c} for the case of strong interference. Note from Lemma 1, the capacity maximization under different scenarios is equivalent to maximizing the strong interference capacity. Therefore, the power allocation problem, which has a constraint on the
outage probability at the primary receiver node (BS), can be formulated as
\begin{align}\label{13}
    &
    \max_{P_1,\ldots,P_K}\,\, \log_2\left(1+\frac{\sum_{k=1}^{K}P_k\sigma_{h_k}^2}{\mathcal{N}_s}\exp(-\kappa)\right),
    \nonumber\\
    &\text{s.t.}\,\,\,
    1- e^{-\frac{\gamma_{\text{th}}\,\mathcal{N}_p}{P_0\sigma_{g_p}^2}}\prod_{k=1}^{K}
    \left(1+\frac{P_k\,\sigma_{g_k}^2}{P_0\,\sigma_{g_p}^2}
    \gamma_{\text{th}}\right)^{-1}\leq\rho_{m},
    \,\,\,\,P_{k}\geq 0,\,\,\text{for}\,\,k=1,\ldots,K.
\end{align}
The objective function in \eqref{13} is a concave function of the
power allocation $P_k$, $k=1,\ldots ,K$, parameters.
Thus, for the convexity of the problem in \eqref{13}, the constraint set $D_f$ must be a convex set. The first constraint in
\eqref{13} is
\begin{equation}\label{14}
    f\left(\{P_{k}\}_{k=1}^{K}\right)=1- e^{-\frac{\gamma_{\text{th}}\,\mathcal{N}_p}{P_0\sigma_{g_p}^2}}\prod_{k=1}^{K}
    \left(1+\frac{P_k\,\sigma_{g_k}^2}{P_0\,\sigma_{g_p}^2}
    \gamma_{\text{th}}\right)^{\!\!-1}\!\!\!-\!\rho_{m},
\end{equation}
with $D_f=\left\{P_{k}\in(0,\infty),
\mid f\left(\{P_{k}\}_{k=1}^{K}\right)\leq0\right\}$, $f:D_f\longrightarrow \mathds{R}$.
Although $f\left(\{P_{k}\}_{k=1}^{K}\right)$ is a convex function of the primary user power $P_0$, it is a concave function of the secondary transmit powers $P_k$, $k=1,\ldots,K$.
Hence, $D_f$ is not a convex set, and thus, this makes the problem nonconvex.

Since the KKT conditions are still valid for non-convex problems, but may lead to a local optimum, in the following, we propose an iterative algorithm based on the KKT conditions. We also solve it through the use the well-established
interior point methods~\cite{boy04}.

The Lagrangian of the problem stated in \eqref{13} is
\begin{equation}\label{16}
    L(\{P_{k}\}_{k=1}^{K})=-\log_2\left(1+\frac{\sum_{k=1}^{K}P_k\sigma_{h_k}^2}{\mathcal{N}_s\,e^{\kappa}}\right)+\lambda
f(\{P_{k}\}_{k=1}^{K}).
\end{equation}
For secondary users $i=1,\ldots,K$ with nonzero transmitter powers, the
KKT conditions are
\begin{align}\label{17}
    &\frac{\partial}{\partial P_{i}}L(\{P_{i}\}_{i=1}^{K})=\frac{-\log_2\!e\,\,\alpha_i}{1+\sum_{k=1}^{K}P_k\alpha_k}+\lambda
\frac{\zeta \beta_i (1+P_i \beta_i)^{-1}}{\prod_{k=1}^{K}(1+P_{k}\beta_k)}
=0, 
\end{align}
\begin{align}\label{17b}
\lambda f(\{P_{k}\}_{k=1}^{K})=0, \,\,\,\,\,\,\,\lambda\geq0,\,\,\,\,\,\,\,f(\{P_{k}\}_{k=1}^{K})\leq0,
\end{align}
where $\alpha_i=\frac{\sigma_{h_i}^2}{\mathcal{N}_s\,e^{\kappa}}$, $\beta_i=\frac{\sigma_{g_i}^2\gamma_{\text{th}}}{P_0\,\sigma_{g_p}^2}$, and $\zeta=e^{-\frac{\gamma_{\text{th}}\,\mathcal{N}_p}{P_0\sigma_{g_p}^2}}$.
Since assuming Lagrange multiplier $\lambda=0$ contradicts the equalities in \eqref{17},
we have always
$f(\{P_{k}\}_{k=1}^{K})=0$. Hence, the problem in \eqref{13} can be reduced to
\begin{align}\label{13m}
    &
    \max_{P_1,\ldots,P_K}\,\, \log_2\left(1+\frac{\sum_{k=1}^{K}P_k\sigma_{h_k}^2}{\mathcal{N}_s}\exp(-\kappa)\right)\!\!,
    \,\,\text{s.t.}\,
    -f(\{P_{k}\}_{k=1}^{K})=0,
    \,\,\,\,P_{k}\geq 0,\,\,\text{for}\,\,k=1,\ldots,K.
\end{align}
where $-f(\{P_{k}\}_{k=1}^{K})$ is a convex function in the feasible set of the power coefficients $P_k$, $k=1,\ldots,K$. Therefore, the problem in \eqref{13m} is a convex optimization, and thus, solving the KKT conditions leads to a global optimum solution \cite[pp. 243]{boy04}.
From \eqref{14}, we have
\begin{equation}\label{20h}
    \prod_{k=1}^{K}(1+P_{k}\beta_k)=(1-\rho_m)^{-1}\zeta.
\end{equation}
Combining \eqref{17} and \eqref{20h}, we can
find the Lagrange multiplier as
\begin{equation}\label{21}
    \lambda=\frac{1+P_i\beta_i}{1+\sum_{k=1}^{K}P_k\alpha_k}\log_2\!(e)\,(1-\rho_m)^{-1}\alpha_i\beta_i^{-1},
\end{equation}
for $i=1,\ldots,K$. From \eqref{21}, power coefficients $P_{j}$, $j=2,\ldots,K$, can be represented in terms of $P_{1}$ as
\begin{equation}\label{21b}
    P_{j}=\frac{1}{\beta_j}\left[\frac{\beta_j\,\alpha_1}{\beta_1\alpha_j}(1+P_1\beta_1) -1 \right]\triangleq
    F_j(P_{1}),
\end{equation}
for $j=2,\ldots,K$. Substituting $P_{j}$ from \eqref{21b} into \eqref{20h}, we can find
$P_{1}$ from the following nonlinear equation:
\begin{equation}\label{20u}
    P_{1}=\left[(1\!-\!\rho_m)^{\!-1}\zeta \prod_{j=2}^{K}(1\!+F_j(P_{1})\beta_j)^{\!-1}\!-\!1\!\right] \beta_1^{-1}\triangleq
    G(P_{1}).
\end{equation}
Then, $P_{j}$, $j=2,\ldots,K$, can be found using \eqref{21b}.

\begin{corollary}
The ergodic capacity maximizing power allocation, when OIC is used, is same as the power allocation coefficients given in \eqref{21b} and \eqref{20u}.
\end{corollary}
\begin{proof}
The proof is followed by using Lemma 1 and the problem formulation in \eqref{13}.
\end{proof}

\emph{
Finding the transmit power limits}: From \eqref{21b}, we can find the maximum allowable power transmitted by each secondary MS. By transmitting the whole power budget from the first node we have $F_j(P_{1})=0$, $j=2,\ldots,K$, and the corresponding transmit power becomes
\begin{equation}\label{20w}
    P_{\max}^i=\left[(1\!-\!\rho_m)^{\!-1}\zeta -1\right] \beta_1^{-1}.
\end{equation}

Moreover, for initial guess about the optimum point, from \eqref{21b} and by the fact that $F_j(P_{1})$ is an increasing function of $P_1$, we can find the minimum value of the transmit power operating point. Since $P_{j}\geq0$, $j=2,\ldots,K$, from \eqref{21b}, we can find that $P_1\geq P_{\min}^i$ where
\begin{align}\label{20w2}
    P_{\min}^i&=\max_{j=2,\ldots,K}\left\{\left[\frac{\alpha_j}{\beta_j\alpha_1}-\frac{1}{\beta_1}\right]^{+}\right\}
=\left[\max_{j=2,\ldots,K}\left\{\frac{\sigma_{h_j}^2}{\sigma_{g_j}^2}\right\}\frac{P_0\,\sigma_{g_p}^2}{\sigma_{h_1}^2\gamma_{\text{th}}}
    -\frac{P_0\,\sigma_{g_p}^2}{\sigma_{g_1}^2\gamma_{\text{th}}}\right]^{+}.
\end{align}
where $[x]^+$ denotes $[x]^+=\max\{0,x\}$.

\emph{
Cognitive system operation condition}:
Since $P_{\max}^i$ should be positive, the condition that cognitive system can co-exists with primary system can be found from \eqref{20w} as
    $1-\rho_m<\zeta$.
By replacing $\zeta$ and $\gamma_{\text{th}}$ with the system parameter, the outage probability margin should satisfy the following condition:
\begin{equation}\label{20m2}
    \rho_m>1-e^{-\frac{(2^{R_p}-1)\,\mathcal{N}_p}{P_0\sigma_{g_p}^2}}\triangleq \rho_0.
\end{equation}
If \eqref{20m2} is not satisfied, the cognitive system should be turned off to not interfere the primary system.
Note that in \eqref{20m2}, $\rho_0$ is basically the amount of outage probability of the primary system in absence of cognitive radios.

\emph{
Recursive Power Allocation Algorithm}:
In Table I, we show an iterative algorithm to numerically find the optimum power allocation. First, we set the initial transmit power $P_{1}$ to a random value in the range of \eqref{20w} and \eqref{20w2}. Then, in the iterative power updating phase, we use equations in \eqref{21b} and \eqref{20u}. Note that from \eqref{20u}, the boundary condition in \eqref{20h} is satisfied in all the iterations. Moreover, since this iterative algorithm is obtained from solving KKT conditions and the fact that a convex optimization problem has a single optimum point, this algorithm converges to the optimal point.

\begin{table}[e]
  \centering
  \caption{
    Maximum rate power allocation of secondary cognitive network with outage constraint at the primary user}
    \begin{tabular}{l}
      \hline
      \emph{Initialization:} \vspace*{.5em}\\
    \hspace*{2em} Initialize $P_1$ from the interval $P_1\in(P_{\min}^i,P_{\max}^i)$ where $P_{\min}^i$\\
     \hspace*{2em} and $P_{\max}^i$ are obtained in \eqref{20w} and \eqref{20w2}, respectively.\vspace*{.5em} \\
    \emph{Recursion:} \vspace*{.5em}\\
    \hspace*{2em}Set $P_j=[F_j(P_{1})]^+$ for $j=2,\ldots,K$, where $F_j(P_{1})$ is given by \eqref{21b}. \vspace*{.5em}\\
    \hspace*{2em}Find $P_1^{\text{new}}=[G(P_{1})]^+$ where $P_1^{\text{new}}$ is the updated version of $P_1$ and\\ \hspace*{2em}$G(P_{1})$ is given by \eqref{20u}.  \vspace*{.5em}\\
    \hspace*{2em}Repeat the recursion until the desired accuracy is
    reached.\vspace*{.4em}\\
      \hline
    \end{tabular}
\end{table}

\emph{
Power Allocation with Power Constraint}:
Now, we consider the case that there is a power constraint in each secondary user, i.e., $P_k\leq P_k^{\max}$ where $P_k^{\max}$ is the maximum power budget of cognitive user $k$. Thus, the optimization problem in \eqref{13} can be rewritten as
\begin{align}\label{13r}
    &
    \max_{P_1,\ldots,P_K}\,\, \log_2\left(1+\frac{\sum_{k=1}^{K}P_k\sigma_{h_k}^2}{\mathcal{N}_s\,e^{\kappa}}\right),
    \nonumber\\
    &\text{s.t.}\,\,\,
    1- e^{-\frac{\gamma_{\text{th}}\,\mathcal{N}_p}{P_0\sigma_{g_p}^2}}\prod_{k=1}^{K}
    \left(1+\frac{P_k\,\sigma_{g_k}^2}{P_0\,\sigma_{g_p}^2}
    \gamma_{\text{th}}\right)^{\!\!-1}\!\!\leq\rho_{m},
    \,\,\,\,0\leq P_{k}\leq P_k^{\max},\,\,\text{for}\,\,k=1,\ldots,K.
\end{align}

In this case, the iterative algorithm given in Table I can be modified as follows: First, we initialize the transmit power with a positive value in the range of $P_{\min}^i$ and $P_1^{\max}-[P_1^{\max}-P_{\max}^i]^{+}$, where $P_{\min}^i$ and $P_{\max}^i$ are defined in \eqref{20w} and \eqref{20w2}, respectively. Then, we can calculate the $P_j$, $j=2,\ldots,K$ as $P_j^{\max}-[P_j^{\max}-F_j(P_1)]^{+}$ where $F_j(P_1)$ is given in \eqref{21b}. The updated value of $P_1$ is computed as $P_1^{\max}-[P_1^{\max}-G(P_1)]^{+}$ where $G(P_1)$ is given in \eqref{20u}. By repeating the procedure stated above, the optimum power coefficients with desired accuracy is achieved.
Table II summarizes the algorithm given above for solving \eqref{13r}.

\begin{table}[e]
  \centering
  \caption{
    Maximum rate power allocation of secondary cognitive network with outage constraint at the primary user and power constraint per user}
    \begin{tabular}{l}
      \hline
      \emph{Initialization:} \vspace*{.5em}\\
    \hspace*{2em} Initialize $P_1$ from the interval $P_1\geq P_{\min}^i$ and\\
     \hspace*{2em} $P_1\leq P_1^{\max}-[P_1^{\max}-P_{\max}^i]^{+}$ where $P_{\min}^i$ and $P_{\max}^i$ are\\
     \hspace*{2em} obtained in \eqref{20w} and \eqref{20w2}, respectively.\vspace*{.5em} \\
    \emph{Recursion:} \vspace*{.5em}\\
    \hspace*{2em}Set $P_j=P_j^{\max}-[P_j^{\max}-F_j(P_1)]^{+}$ for $j=2,\ldots,K$, where \\
    \hspace*{2em} $F_j(P_{1})$ is given by \eqref{21b}. \vspace*{.5em}\\
    \hspace*{2em}Find $P_1^{\text{new}}=P_1^{\max}-[P_1^{\max}-G(P_1)]^{+}$ where $P_1^{\text{new}}$ is the \\ \hspace*{2em} updated version of $P_1$ and $G(P_{1})$ is given in \eqref{20u}.  \vspace*{.5em}\\
    \hspace*{2em}Repeat the recursion until the desired accuracy is
    reached.\vspace*{.4em}\\
      \hline
    \end{tabular}
\end{table}

\subsection{Power Optimization with Known Cognitive MAC Instantaneous CSI at the Secondary Users}
Here, we assume that instantaneous CSI of cognitive multiple access channel gains are available at the secondary users. However, only statistics of the interference channels, i.e., $\sigma_{g_k}^2$, $k=1,\ldots,K$, can be estimated.
We first present the results for the strong primary interference case.
Thus, we consider the instantaneous achievable rate in \eqref{1b} as a performance metric at the cognitive MAC system.
Therefore, the power allocation problem, which has a
required outage probability constraint on the primary BS node, can be
formulated as
\begin{align}\label{131}
    &
    \max_{P_1,\ldots,P_K}\,\, \log_2\left(1+\frac{\sum_{k=1}^{K}P_k |h_k|^2}{\mathcal{N}_s}\right),
    \nonumber\\
    &\text{s.t.}\,\,\,
    1- e^{-\frac{\gamma_{\text{th}}\,\mathcal{N}_p}{P_0\sigma_{g_p}^2}}\prod_{k=1}^{K}
    \left(1+\frac{P_k\,\sigma_{g_k}^2}{P_0\,\sigma_{g_p}^2}
    \gamma_{\text{th}}\right)^{-1}\leq\rho_{m},
    \,\,\,\,P_{k}\geq 0,\,\,\text{for}\,\,k=1,\ldots,K.
\end{align}

\begin{proposition}
The solution for the power allocation values $P_{k}^*$, $k=1,\ldots,K$ in the optimization problem \eqref{131} can be expressed as
\begin{equation}\label{41u}
    P_{1}=\left[(1\!-\!\rho_m)^{\!-1}\zeta \prod_{j=2}^{K}(1\!+\widetilde{F}_j(P_{1})\beta_j)^{\!-1}\!-\!1\!\right] \beta_1^{-1}\triangleq
    \widetilde{G}(P_{1}),
\end{equation}
\begin{equation}\label{41b}
    P_{j}=\frac{1}{\beta_j}\left[\frac{\beta_j\,|h_1|^2}{\beta_1\,|h_j|^2}(1+P_1\beta_1) -1 \right]\triangleq
    \widetilde{F}_j(P_{1}),
\end{equation}
for $j=2,\ldots,K$.
\end{proposition}
\begin{proof}
The proof is similar to the procedure given in Subsection IV-A which lead to \eqref{21b} and \eqref{20u}.
\end{proof}

The iterative algorithm expressed in Table I can be also used for the scenario given in this subsection, where instantaneous CSI of cognitive network is known at the secondary users. But functions ${F}_j(P_{1})$ and ${G}(P_{1})$ are replaced by $\widetilde{F}_j(P_{1})$ and $\widetilde{G}(P_{1})$, respectively, and ${P}_{\min}^i$ in \eqref{20w2} can be rewritten as
\begin{align}\label{20w3}
    \widetilde{P}_{\min}^i&=\left[\max_{j=2,\ldots,K}\left\{\frac{|h_j|^2}{\sigma_{g_j}^2}\right\}
    \frac{P_0\,\sigma_{g_p}^2}{|h_1|^2\gamma_{\text{th}}}
    -\frac{P_0\,\sigma_{g_p}^2}{\sigma_{g_1}^2\gamma_{\text{th}}}\right]^{+}.
\end{align}

\begin{corollary}
The instantaneous capacity maximizing power allocation when OIC is used is the same as the power allocation coefficients given in \eqref{21b} and \eqref{20u}.
\end{corollary}
\begin{proof}
The proof is similar to the proof of Corollary 1.
\end{proof}

\begin{figure}[e]
  \centering
  \vspace{-.3cm}
  \includegraphics[width=\columnwidth]{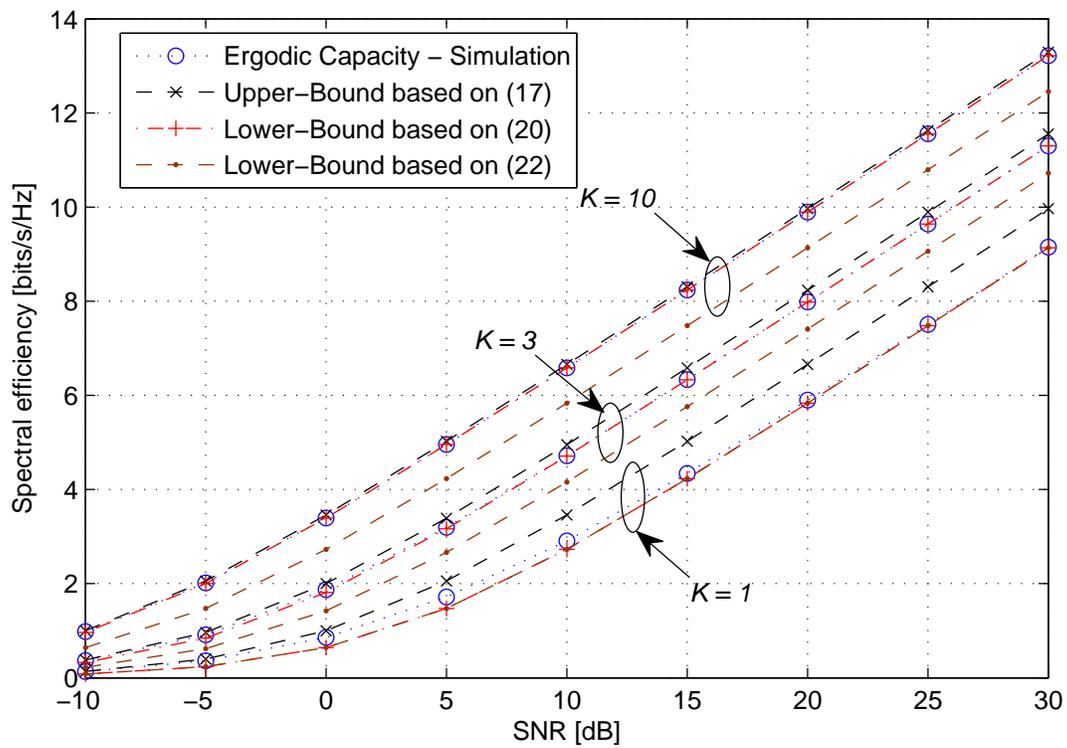}\\
  \vspace{-.3cm}
  \caption{Ergodic sum rate of the secondary multiple access system for one and two users when interference is strong and can be decoded, i.e., clean MAC. Upper and lower bounds are also depicted.
  }\label{fb}
\end{figure}
\begin{figure}[e]
  \centering
  \vspace{-.3cm}
  \includegraphics[width=\columnwidth]{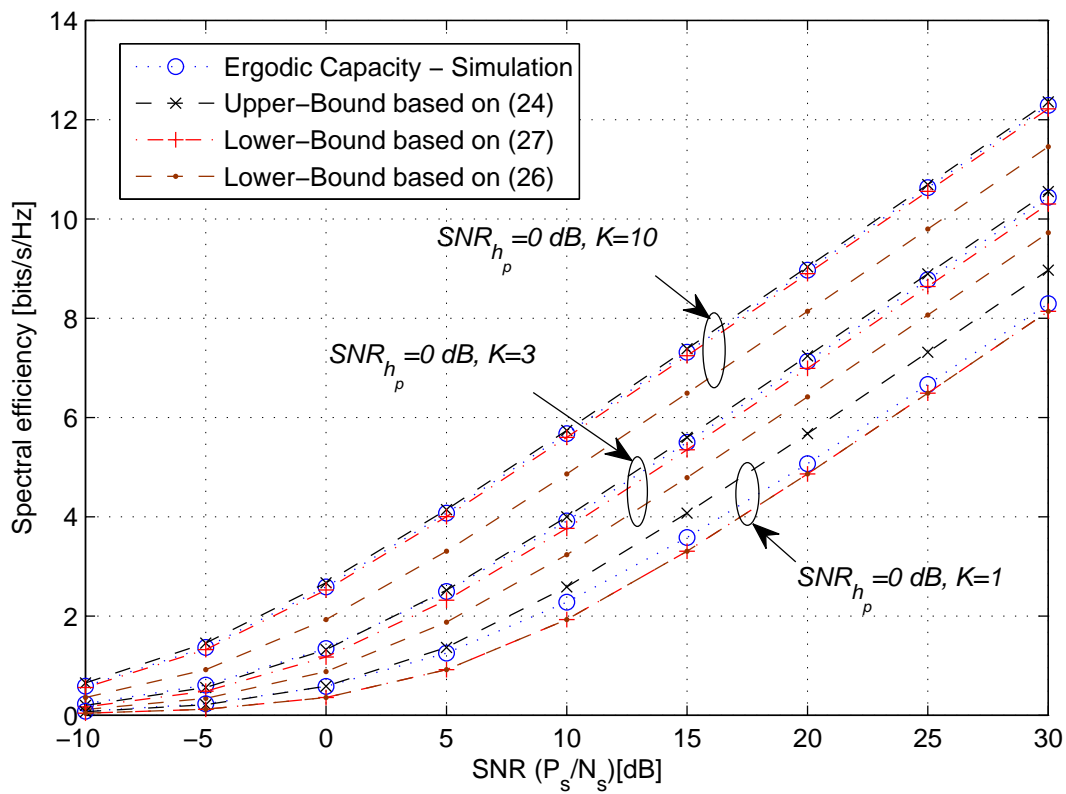}\\
  \vspace{-.3cm}
  \caption{Ergodic sum rate of the secondary multiple access system for one and two users when interference from primary user is treated as noise and $\text{SNR}_{h_p}=\frac{P_0|h_p|^2}{\mathcal{N}_s}=1$. An upper-bound and two approximations are also depicted.
  }\label{fb2}
\end{figure}

\begin{figure}[e]
  \centering
  \vspace{-.3cm}
  \includegraphics[width=\columnwidth]{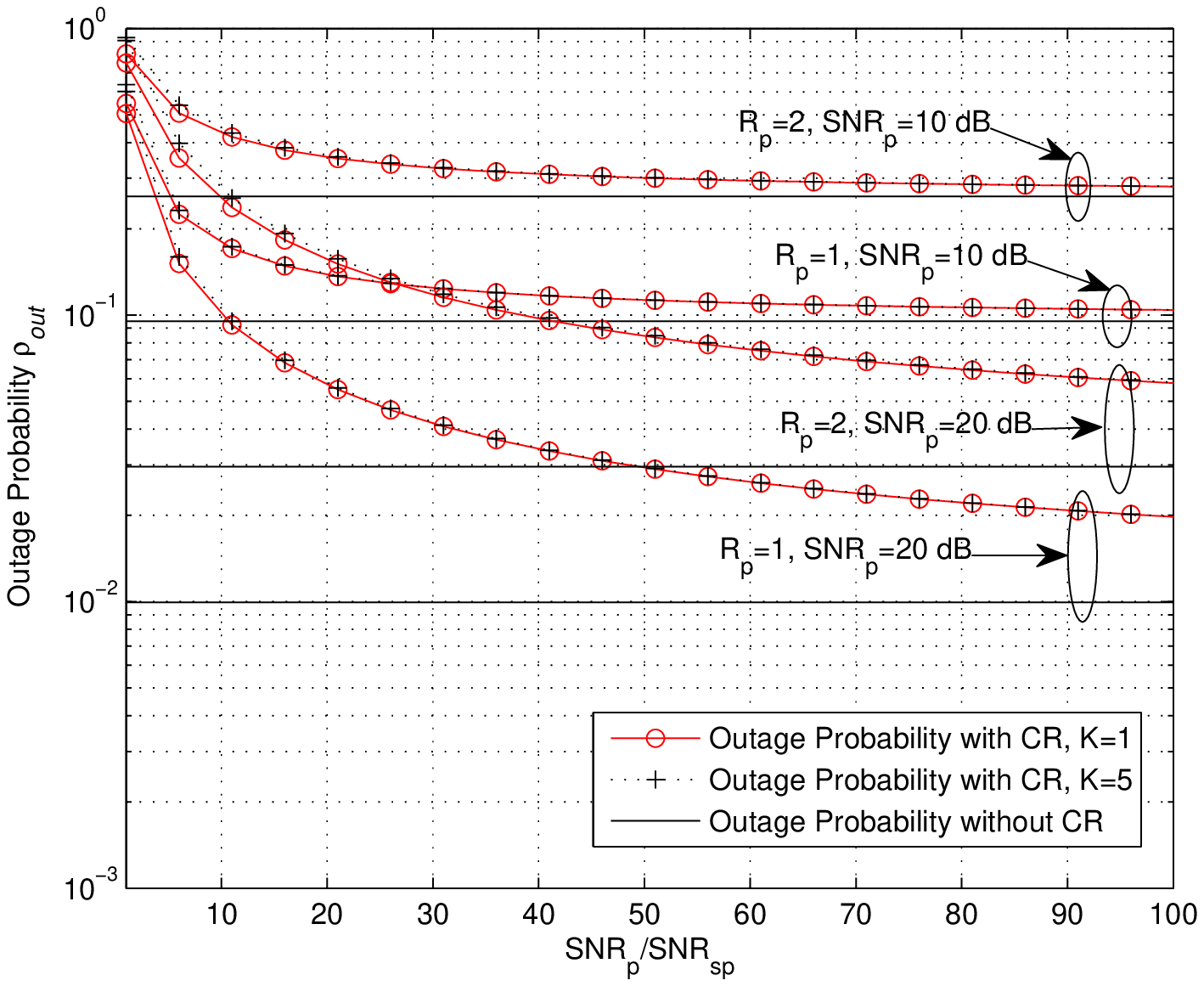}\\
  \vspace{-.3cm}
  \caption{Outage probability in the primary system as a function of the ratio between the average diffuse component of the primary signal and the average SNR of the interfering signal from the secondary at the primary receiver. The systems with different number of users $K$, primary rate $R_p$ and average primary $\text{SNR}_p$ are compared. 
  }\label{fc}
\end{figure}

\vspace{-.2cm}
\section{Numerical Analysis}
In this section, numerical results are provided to demonstrate the usefulness of our analytical results, as well as the effectiveness of the resource allocation algorithms presented in previous sections. We consider a $K$ users secondary system with a common BS.
In all the evaluation scenarios we have assumed that the secondary system multiple access links $h_k$ and interference links $g_k$ are independent Rayleigh distributed with variance $\sigma_{h}^2$ and $\sigma_{g}^2$, respectively.


In Fig.~\ref{fb}, the ergodic rate $\overline{C}_{\text{sum}}$ in \eqref{11} achievable with SIC for $K=1,3$ is depicted. The upper and lower bounds on sum capacity derived in Subsection III-B are also depicted. The horizontal axis is transmit SNR from each secondary user. As it can be seen the upper and lower bounds are tight for both cases of one and three users. From the figure, we observe that the lower-bound based on \eqref{12c} is very close to the capacity. However, the lower bound in \eqref{12c} is only valid for i.i.d. distributed cognitive radio channels $h_k$. In contrast, the lower bound based on \eqref{120c} can be also used for non-i.i.d. distributed links.

In Fig.~\ref{fb2}, the ergodic rate $\overline{C}_{\text{sum}}^{\text{int}}$ in \eqref{12ss} in which interference from the primary node is treated as noise is depicted for $K=1,3$. It is also assumed that the received SNR from the primary transmitter, i.e., $\text{SNR}_{h_p}=\frac{P_0\sigma_{h_p}^2}{\mathcal{N}_s}$ is 0 dB and cannot be decoded at the secondary receiver, and thus, it is treated as noise. The approximation on sum capacity derived in Subsection III-B-3 are also depicted. The horizontal axis is transmit SNR from each secondary user, i.e., $\frac{P_k}{\mathcal{N}_s}$. As it can be seen, the upper bound and approximations are tight for both cases of one and three users. Although the approximation based on \eqref{12x} is not necessarily a lower-bound, it can be seen from simulations that this approximation is a lower bound on the capacity for the two cases demonstrated in Fig.~\ref{fb2}.

Fig.~\ref{fc} considers the outage probability experienced in the primary system as a function of the ratio between the average diffuse primary component and the average interference power received from the secondary transmitters, which is denoted by $\frac{\text{SNR}_p}{\text{SNR}_{sp}}$ where $\text{SNR}_p=\frac{P_0\sigma_{g_p}^2}{\mathcal{N}_p}$ and $\text{SNR}_{sp}=\frac{\sum_{k=1}^K P_k\sigma_{g_k}^2}{\mathcal{N}_p}$. The transmission rate in the primary system $R_p$ is fixed to 1 and 2 bits/channels use and we measure the outage probability in the primary system. From \eqref{10}, it can be seen when the power ratio goes to infinity, the outage probability converges to the case of outage probability without cognitive radio, i.e., $\rho_{0} = 1- e^{-\frac{2^{R_p}-1}{\text{SNR}_p}}$. The curves are shown for different values of $R_p$ and $\text{SNR}_p$, and it can be seen that for a fixed amount of interference from the secondary system, lower primary rate and higher primary SNR reduces the outage probability at the primary node. Another interesting observation from Fig.~\ref{fc} is that the outage margin is more sensitive when $\text{SNR}_p$ is increasing. In other words, the difference between target outage probability $\rho_m$ and $\rho_0$ (the outage probability in absence of cognitive radio) is higher for a larger $\text{SNR}_p$. Another observation is that it is shown that by changing the number of user from $K=1$ user to $K=5$ user, the outage probability is not much varying for all cases depicted in Fig.~\ref{fc}.

\begin{figure}[e]
  \centering
  \vspace{-.3cm}
  \includegraphics[width=\columnwidth]{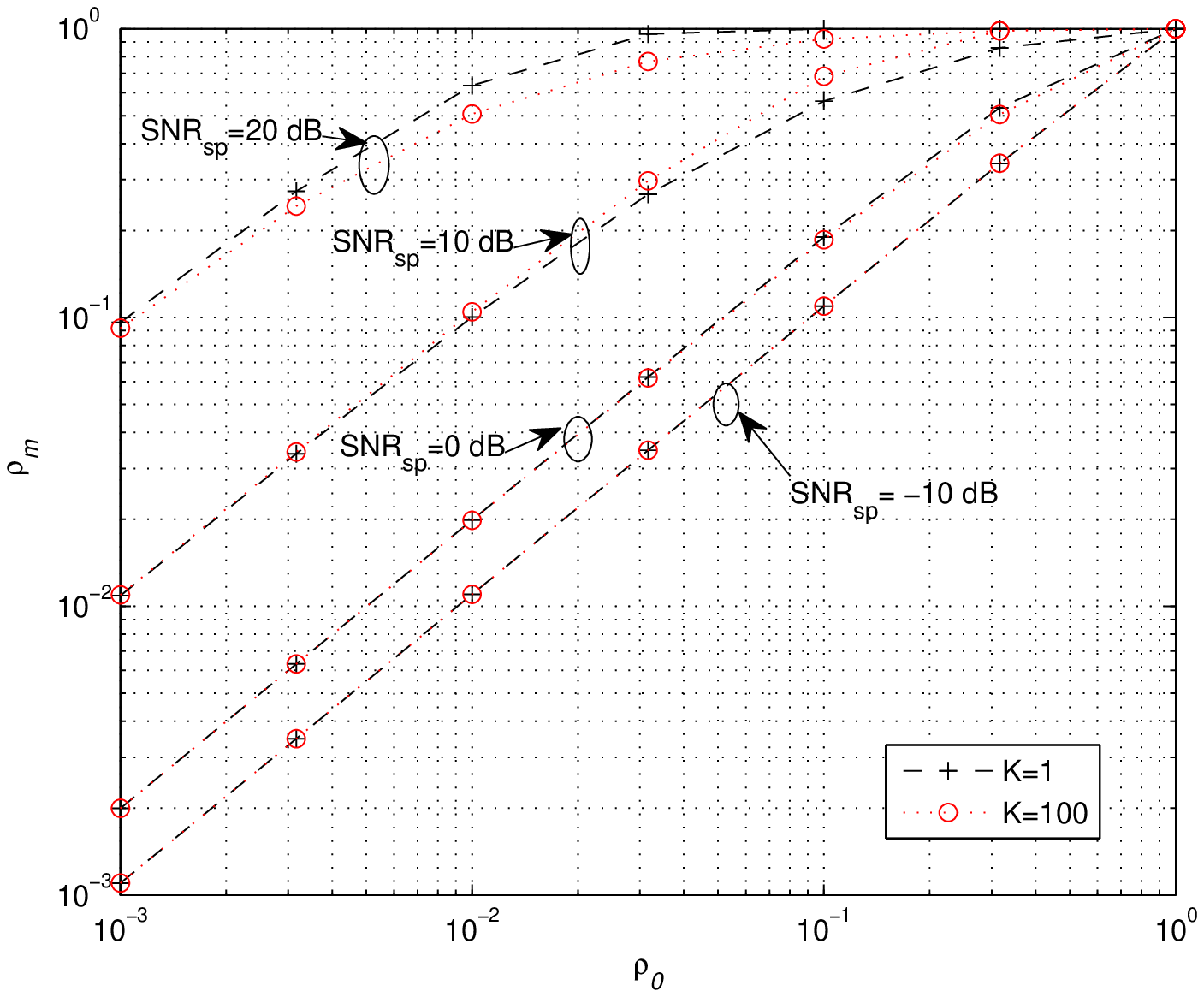}\\
  \vspace{-.3cm}
  \caption{The target outage probability at primary node $\rho_m$ in presence of CR versus the outage probability in absence of CR for different values of average interference SNR at primary receiver and different number of users. 
  }\label{fd}
\end{figure}

In Fig.~\ref{fd}, we compare the target outage probability at primary node $\rho_m$ in the presence of CR versus the outage probability in absence of CR for different values of average interference SNR at primary receiver, i.e., $\text{SNR}_{sp}$, and different number of users $K=1, 100$. It can be seen that as interference parameter $\text{SNR}_{sp}$ goes down, the outage probability gets closer to $\rho_0$. However, for high interference from CR and higher value of $\rho_0$, the outage margin at the primary user becomes too high, and hence, co-existence of primary and secondary is not feasiblec. Moreover, it is also observable that the relationship between $\rho_m$ and $\rho_0$ is not sensitive to the number of users $K$, especially for lower interference powers from secondary nodes.

\begin{figure}[e]
  \centering
  \vspace{-.3cm}
  \includegraphics[width=\columnwidth]{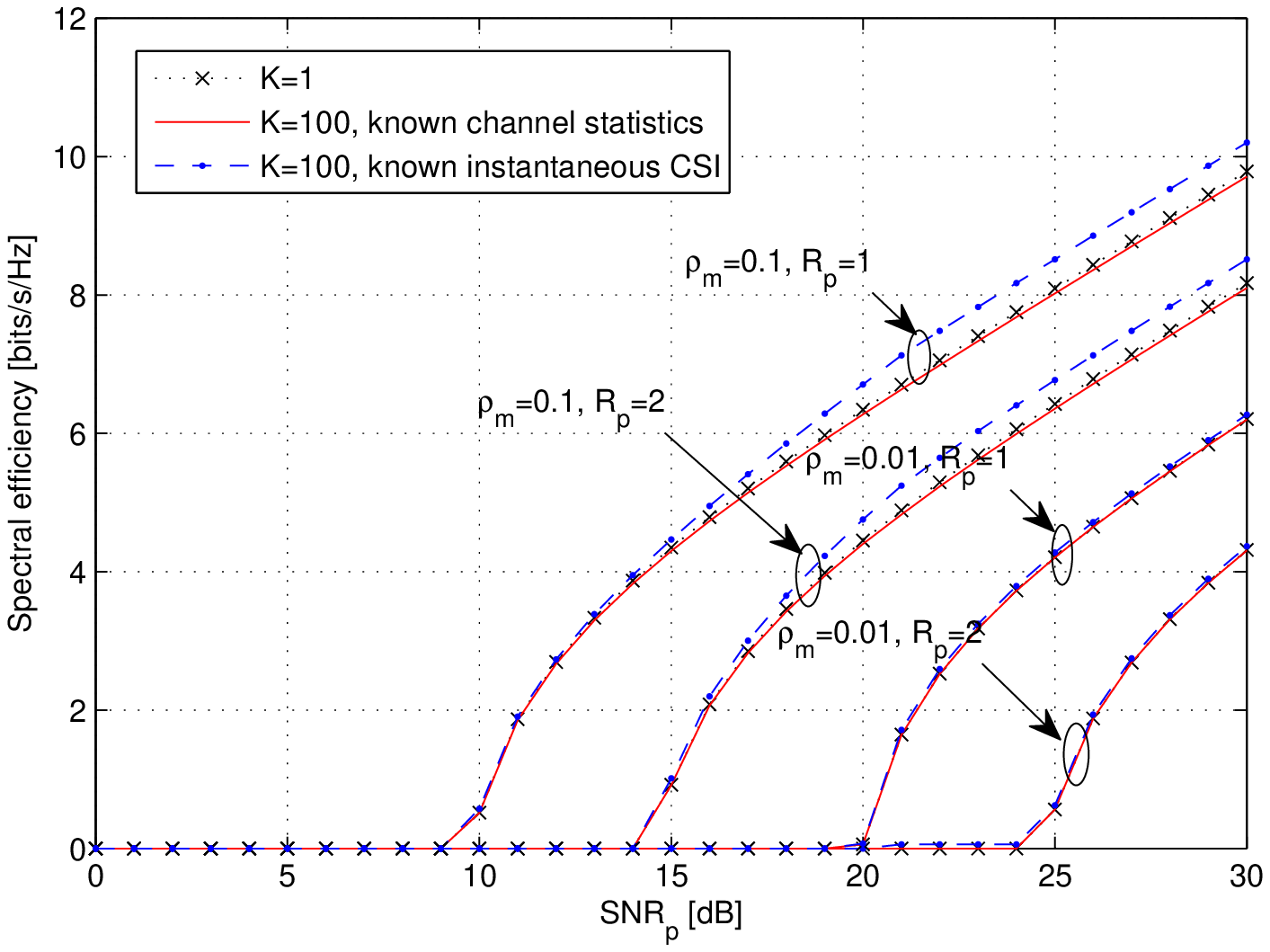}\\
  \vspace{-.3cm}
  \caption{Ergodic capacity of the secondary users as a function of average SNR of the primary system in a network with $K=1,100$ users, outage targets of $\rho_{m}=10^{-1}, 10^{-2}$, and primary rates $R_p=1,2$ bits/s/Hz.
  }\label{fe}
\end{figure}

Finally, Fig.~\ref{fe} shows the achievable sum-rate capacity of the secondary system for different primary outage target, primary rate, and number of users. For calculating the achievable capacity, the maximum allowable power is found using algorithms given in Section IV. We have also assumed that the distance of the secondary users from the primary BS are two times of their distance from secondary BS, i.e., $\frac{\sigma_{h_k}^2}{\sigma_{g_k}^2}=8$ when the path-loss exponent is equal to 3. It can be seen that when the SNR of the primary system is low, the CR system should be turned off. For example, the threshold $\text{SNR}_p$ for operating point of CR is 14 dB when $\rho_m=10^{-2}$ and $R_p=1$ bits/s/HZ.
Furthermore, from Fig.~\ref{fe} it is observed that for higher target outage $\rho_m$ and lower primary rate $R_p$, the secondary capacity is increased. In this numerical example, we have also observed that when the outage probability
$\rho_{m}=10^{-2}$ is required at primary receiver, and $\text{SNR}_p=25$ dB, by decreasing $R_p$ from 2 to 1 bits/s/Hz, capacity of secondary system is increased around 3.5 bits/s/Hz.
Now, we study the asymptotic behavior of the curves in Fig.~\ref{fe}. Assuming $\sigma_{h_k}^2=\sigma_{h}^2$ and $\sigma_{g_k}^2=\sigma_{g}^2$, for $k=1,\ldots,K$, a closed-form solution for the transmit power of each secondary user can be found from \eqref{21b} and \eqref{20u} as
\begin{equation}\label{71uu}
    P_{k}^*=
    \frac{P_0\,\sigma_{g_p}^2}{\sigma_{g}^2\gamma_{\text{th}}}
    \left[e^{-\frac{\gamma_{\text{th}}\,\mathcal{N}_p}{K\,P_0\sigma_{g_p}^2}}(1-\rho_m)^{-1/K}-1\right] .
\end{equation}
Thus, form \eqref{120c} and \eqref{71uu}, the slope of the ergocic capacity in Fig.~\ref{fe} in high SNR scenario is given by
\begin{align}\label{120cc}
    \lim_{\text{SNR}_p\rightarrow \infty}\frac{\overline{C}_{\text{sum}}}{10\log_{\!10}(\text{SNR}_p)}&=
    \frac{\log_2\left(10\right)}{10}\approx 0.33.
\end{align}
Finally, we again see that the ergodic capacity of the secondary system is not sensitive to the number of users. Nonetheless, for the case of $\rho_{m}=10^{-1}$, a single user cognitive network achieves slightly higher capacity gain than a network with $K=100$ users. In addition, since increasing the number of users does not have much effect on the sum-rate capacity, it can be inferred that the proposed system can achieve considerable gain in spectrum efficiency compared to orthogonal transmission strategies.

\vspace{-.2cm}
\section{Conclusion}
We have considered communication scenarios in which the
secondary (cognitive) uplink users are allowed to transmit along with
the transmissions in the primary system, not violating the
target outage performance in the primary system. This paper formulated the power allocation problem to maximize the sum-rate of cognitive radio users on Gaussian MAC when there is outage constraint at the primary user. We proposed efficient and simple solutions for the power control. The secondary
transmitters can guarantee the outage probability
for a primary terminal by appropriate assigning the
transmit power.
A simple closed form expression for the outage probability at the primary user was derived. Various tight lower and upper bounds were found for the ergodic sum-rate capacity of the secondary system.
We have
also investigated that the secondary users should apply OIC and cancel the interference from the primary system
whenever such opportunity is created by (a) selection of the
data rate in the primary system and (b) the link quality between the primary transmitter and the secondary receiver. We devised a method for obtaining a maximal achievable rate in the uplink secondary system whenever the primary signal is decodable. The numerical results confirmed that the proposed schemes can bring rate gains in the CR systems.


%


\bibliographystyle{ieeetr}
\bibliography{references}
\end{document}